\documentclass[slac_one]{revtex4}
\usepackage{graphicx}
\usepackage{fancyhdr}
\begin{document}

\title{Commissioning of the ATLAS Inner Tracking Detectors} %% Paper title goes here


\author{F. Martin}
\affiliation{University of Pennsylvania, Philadelphia, PA 19104, USA}
\author{On behalf of the ATLAS Inner Detector Collaboration}
%\affiliation{FNAL, Batavia, IL 60510, USA}

\begin{abstract}
ATLAS (A Toroidal LHC Apparatus) is one of the experiments installed on the CERN Large Hadron Collider ({\bf LHC}), where first collisions are expected in September this year at a reduced center of mass energy of 10 TeV, for the collider commissioning. In this paper, a short description of the Transition Radiation Tracker, of the SemiConductor Tracker and of the Pixel detector, which all together form the ATLAS inner detector, is given, and we discuss their first commissioning results. We conclude with the plans until the {\bf LHC} start-up.  
\end{abstract}

\maketitle

\thispagestyle{fancy}


\section{THE ATLAS INNER DETECTOR} % Section title should be in all capitals.
The ATLAS experiment is described in \cite{ATLASpaper}. The Transition Radiation Tracker ({\bf TRT}), the Semiconductor Tracker ({\bf SCT}) and the {\bf Pixel} detector must provide track identification for $\eta$ ($\eta = -ln (tg (\frac{\theta}{2}))$, where $\theta$ is the polar angle defined {\it wrt} the $z$ axis, along the beam) lower than 2.5 and transverse momentum greater than 500 MeV, with a momentum resolution of $\frac{\sigma_{P_T}}{P_T} = 0.05\% P_T \oplus 1\%$. A good vertex reconstruction is also requested, as well as electron identification capabilities provided by the transition radiation in the {\bf TRT}. The inner detector ({\bf ID}) is 6.2 m long, has a diameter of 2.1 m and is centered on the beam axis. It is located in the ATLAS solenoid, which provides a 2 T magnetic field for charge identification. These detectors have to support a high occupancy at the nominal luminosity of 10$^{34}$ cm$^{-2}$ s$^{-1}$ and a high radiation level, particularly for the first {\bf Pixel} layer. This layer has to be replaced after $\sim$ three years of operation, due to radiation damages. This high radiation environment is also a reason for operating the two silicon detectors at a low temperature (-5 $^o$C to -10 $^o$C). \\
The {\bf TRT} is the furthest of the interaction point, starting at a radius of $\sim$ 55 cm and ending at 1 m. The sensitive elements are small proportional drift tubes called straws \cite{TRTstraw}. The cathode is the carbon-fiber straw wall at a potential of $\sim$ -1.5 kV, the anode is a 30  $\mu m$ gold plated tungsten wire centered in the straw. These 4 mm diameter straws are filled with a Xe (70\%), CO$_2$ (27\%), O$_2$ (3\%) gas mixture during the normal operations. Due to the high Xenon cost, the {\bf TRT} is operated with an Ar (70\%), CO$_2$ (30\%) gas mixture during the commissioning phase. The {\bf TRT} consists off one barrel \cite{TRTbarrel}, and two end-caps \cite{TRTendcap}. The barrel is made of 96 modules, with 1.5 m long axial straws, splitted at their middle to reduce the occupancy. Each straw side is readout independently by the front-end electronics \cite{TRTelectronics}, giving  105088 readout channels. The barrel straws are embedded in a polypropylene foam to provide the transition radiation for electron identification. The end-caps are made of 20 wheels each. Each wheel is made of 8 planes of 39 cm long radial straws. There are two types of wheel, the spacing between the straws being increased with $\eta$, to get a roughly constant number of hits on track with $\eta$. The straw planes are separated by polypropylene foil, again for the electron identification through transition radiation. The total number of channels for the two end-caps is 245760. This is the Transition Radiation Tracker with the highest counting rate and total accumulate charge (10 C/cm over the expected 10 years of the ATLAS experiment) ever built. \\
The {\bf SCT} consists on one barrel with four layers of sensors, at radius 299, 371, 443, 514 mm, and two end-caps \cite{SCTendcap}, with 9 disks of sensors. Sensors are made of double sided (with a stereo angle of 40 mRad) micro-strips of pitch 80 $\mu m$ in the barrel, 57 to 90 $\mu m$ in the end-caps, for a 285$\pm$15 $\mu m$ thickness. There are 2112 barrel modules of 2x768 12 cm long active strips, and 1976 end-cap modules of 4 different types, depending on the position on the disk. The initial operating voltage is 150 V, but it can be increased to 250-450 V after 10 years of operation. There are 6.3 millions readout channels, the noise occupancy per channel being less than 5 10$^{-4}$ at a threshold of 1 fC. \\
The {\bf Pixel} detector is made of 3 barrel layers, at radius 5, 9, and 12 cm, and of two end-caps with 3 disks each. The pixel have a size of 50x400 mm$^2$, for a 250 $\mu m$ thickness. The 80.4 millions readout channels are distributed over 1744 modules.\\
Both {\bf Pixel} and {\bf SCT} used a C$_3$F$_8$ evaporating cooling (see section \ref{Pixel}). 
%Website\footnote{http://www-conf.slac.stanford.edu/lcws05/author$_-$instructions.htm}
\section{THE ATLAS INNER DETECTOR COMMISSIONING}
After many years of assembly and tests on surface \cite{TRTconstruction}, the various elements of the {\bf ID} have been assembled in the ATLAS experimental cavern. This assembly phase is now ended and the detector is closed. Commissioning of the detectors with cosmic rays now takes place. Four different trigger systems are used, with various trigger rates: scintillators installed on the top of ATLAS ($\sim$ 1Hz), trigger towers from the Hadronic Tile Calorimeter \cite{ATLASpaper},\cite{TileCalTrigger} (smaller than 1Hz), level 1 trigger (based on muon chambers and ``coarse'' calorimeter informations, with a trigger delay of 2-3 $\mu s$ at a rate of a few Hz), muon chambers (Resistive Plate Chambers ({\bf RPC}), $\sim$ 100 Hz). A variable (from mHz to 150 kHz) random trigger is also used to make some acquisition rate tests.
 
\subsection{TRT commissioning results}
\begin{figure*}[t]
\begin{minipage}[c]{.46\linewidth}
\includegraphics[width=60mm]{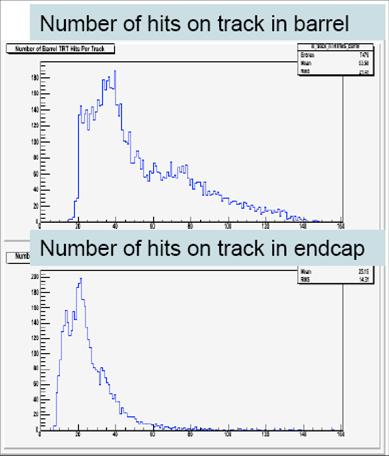} 
\end{minipage}\hfill
\begin{minipage}[c]{.46\linewidth}
\includegraphics[width=60mm]{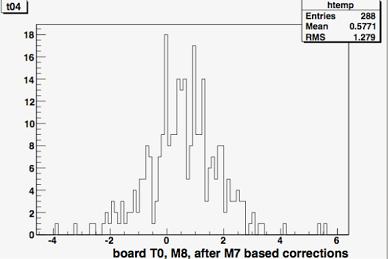}
\end{minipage}\hfill
\caption{Number of hits on track for the 7476 barrel and 3921 end-cap TRT tracks, registered during the commisionning run of July 15$^{th}$-July 21$^{st}$ (left). Trailing edge spread obtained using the data to fit the $T_0$ per front-end board, after correction for the chip to chip spread (right), for the same run time period. The RMS is 1.3 ns.} \label{TRTFig}
\end{figure*}

\begin{figure*}[t]
\includegraphics[width=60mm]{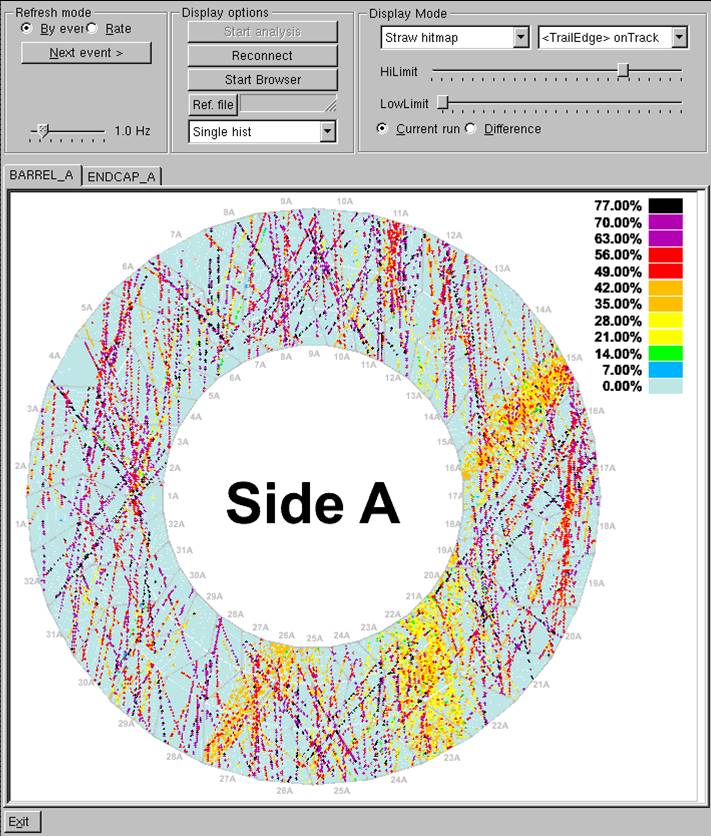} 
\caption{View of the cosmic tracks registered in the {\bf TRT} barrel during the ``M8'' commissioning run. The view is obtained with one of the monitoring tool called ``TRTViewer''.} \label{TRTtrack}
\end{figure*}

Before any data taking, the timing parameters of the detectors have to be tuned. We take the {\bf TRT} as an example. To know the track position {\it wrt} the straw wire, both the leading edge and the trailing edge of the signal are recorded. The trailing edge is fixed by the straw size and the electron drift speed, as there are always some clusters produced close to the straw wall when the track crosses the straw. The arrival time of the leading edge is fixed by the closest approach distance to the straw wire. The maximum drift time is 48 ns, to which the minimum time over threshold (7.5 ns), the chip to chip spread (5 ns) and two empty bins to define the leading and trailing edge (6.25 ns) have to be added. The resulting signal time of $\sim$ 70 ns has to be fit in the readout window of 75 ns, which requires a carreful timing of the {\bf TRT} (see \cite{MH} for more details). As all the clocks on the front-end electronics are derived from the same 40.08 MHz {\bf LHC} clock, any clock phase difference must come from different propagation delays inside the cables or in the electronics. Most of the different types of cables have been measured for signal propagation with a precision of $\pm$3 ns (for some cables the propagation time has been calculated from the known cable properties and from the construction length). Knowing the cable propagation times, it is easy to adjust the three {\bf TRT} timing parameters: the Trigger Delay Mode (coarse timing, by 25 ns step) the clock phase and the data phase on the front-end electronics (fine delays, by 0.5 ns step). This allows to set a preliminary timing, with a trailing edge spread of $\pm$ 5 ns around the average, before applying any corrections to take into account the time of flight difference between the top and bottom modules. This is the order of magnitude of what is expected from the chip to chip spread on the front-end boards \cite{NoteBen}. This preliminary timing can be improved, by fitting the particle crossing time T$_0$  per front-end board, using the data. It is expected to reach a 1 ns precision in the timing after this procedure (present result is $\sim$ 1.3 ns, see fig. \ref{TRTFig}).\\
The low thresholds have also to be set in such a way that the occupancy (the rate of leading edges due to noise times the 75 ns readout window) is uniform across the {\bf TRT}. The variation of the occupancy ({\it O}) with the threshold ({\it T}) can be parametrized by $ T = \alpha \ erf^{-1} (\beta O) + \gamma $. A fit is performed on a sample of noise-only data, and the obtained parameters are used, knowing a measured ({\it O$_m$},{\it T$_m$}) and a target ({\it O$_t$}), to compute the corresponding threshold ({\it T$_t$}). Typically, two iterations are enough to obtain the thresholds corresponding  to a uniform occupancy of 1\%, starting from a non-tuned detector (occupancy between 0 to 10\%) \cite{NoteRyan}.\\ 
The other achievements of the {\bf TRT} commissioning can be summarized, and more details are given in \cite{JD}:
\begin{itemize}
\item The number of suspicious channels after the installation is of the order of 2\%. True dead channels will be identified with the data. 
\item The {\bf TRT} is completely equipped with readout on one side (side ``A''), and partially on the second side (side ``C''). Successful cosmic runs have been taken and more than 11000 tracks have been reconstructed on side ``A'', both in the barrel and the end-cap, during the last one week long cosmic run (see fig. \ref{TRTFig}). The side ``C'' is only partially equipped because of a late delivery of the Read Out Drivers ({\bf RODs}). 
\item The achieved data acquisition rate is 20kHz without 0 suppression, and is limited by the Level 1 trigger. With the 0 suppression, the acquisition rate is 80 kHz. This implies also that the 0 suppression algorithm has been validated.
\item The Detector Control System ({\bf DCS}) for the {\bf TRT} is almost complete and is proved to behave as expected.
\item The monitoring tools have been widely experienced, and cross-checked against each other: see figure \ref{TRTtrack} for a view of registered cosmic tracks in the {\bf TRT} barrel.
\item The straw cooling (gaseous CO$_2$) and the mono-phase electronics cooling (liquid $C_6F_{14}$) have been operated during long periods, without any problems.
\item The detector is under High Voltage $\sim$20\% of the time since months, again without any problems.
\item The active gas system has started to work with the detector.
\item Procedure for the Low Voltage tuning has been tested (this is important, as it strongly affects the thresholds). Algorithms to interact with the {\bf DCS} have been set-up and tested.\end{itemize}

\subsection{SCT commissioning results}
\begin{figure*}[t]
\begin{minipage}[c]{.46\linewidth}
\includegraphics[width=60mm]{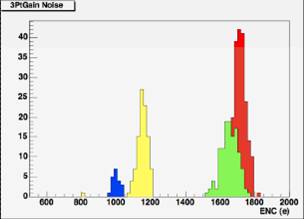} 
\end{minipage}\hfill
\begin{minipage}[c]{.46\linewidth}
\includegraphics[width=60mm]{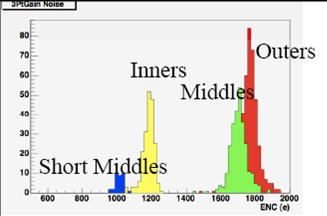}
\end{minipage}\hfill
\caption{Result of the SCT noise scan at room temperature (25$^o$C). The four peaks correspond to the four SCT end-cap module types. A good noise level of 1002, 1149, 1653 and 1707 electrons is measured on the ``A'' side (left); while a noise of 1013, 1182, 1704 and 1767 electrons is measured on the ``C'' side (right).} \label{SCTFig}
\end{figure*}

The {\bf SCT} commissioning period has been interrupted by a cooling plant failure (see section \ref{Pixel}) after 30 days. Nevertheless, a large amount of results have been collected. The full detector has been checked after the installation and the cabling: 99.7\% of the barrel channels are behaving properly, and all the 44 cooling loops \cite{SiliconCooling} are operational. Three cooling loops have to be operated at a lower temperature than the one used during the commissioning, to minimize the stress on the seals. In the endcaps, 99.1\% of the channels are functional, the main source of dead channels (0.65\%) being a leaky cooling loop which prevents 13 modules from running. The other 71 cooling loops are operational, with some tolerable leak on one of them. There are still 5\% of the data links which report some occasional errors (like event or bunch crossing identifier mismatch), but these will be reduced when the timing and the thresholds would have been tuned in the future commissioning. The noise performances have been measured, and have shown good results (see fig. \ref{SCTFig}). The {\bf SCT} has participated in one of the ``global'' ATLAS commissioning run, where the different parts of the detector run together. The complete barrel, except the 3 cooling loops mentioned above, has run in a very stable way, both from the point of view of the detector and of the cooling/heaters system. The tuning of the timing parameter during the cosmic run was performed in few hours (which is reasonable given the low trigger rate of $\sim$0.5 Hz) and the acquisition was stable up to a trigger rate of $\sim$8 kHz. No limit tests have been performed, but with this configuration a rate of 40 kHz should be reachable. The monitoring was also tested and behave quite well, correlated tracks between the {\bf SCT}, the {\bf TRT} and the muon {\bf RPC's} are reconstructed and monitored. The {\bf DCS} has been also widely exercised and proved to be robust. The {\bf SCT} track resolution has been measured to be 65 $\mu m$, which compares well with the previous measurements on the surface, and the combined {\bf TRT} plus {\bf SCT} preliminary resolution is better than 300 $\mu m$. More on this subject can be found in \cite{Muge}.\\
The {\bf SCT} has implemented a {\bf ROD} simulator, in order to join the data acquisition chain in the various ATLAS tests, even if the detector can not be powered due to the cooling plant failure already mentioned. Also, the collaboration has set up on the surface a complete barrel sector, in order to train the shifters and also to test the future software and firmware developments.
 
\subsection{Pixel commissioning results \label{Pixel}} 
\begin{figure*}[t]
\begin{minipage}[c]{.46\linewidth}
\includegraphics[width=60mm]{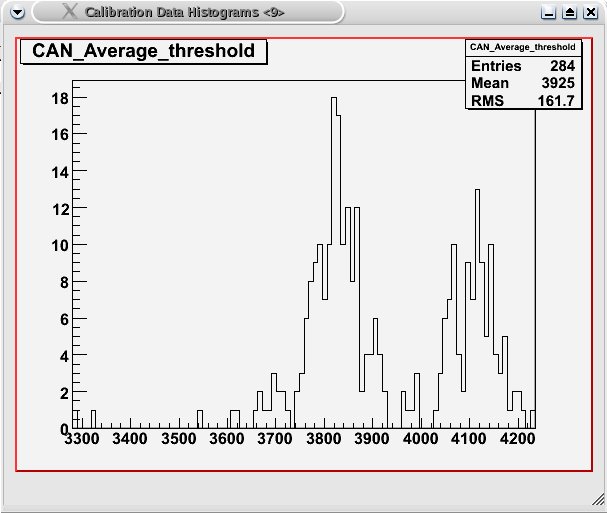} 
\end{minipage}\hfill
\begin{minipage}[c]{.46\linewidth}
\includegraphics[width=60mm]{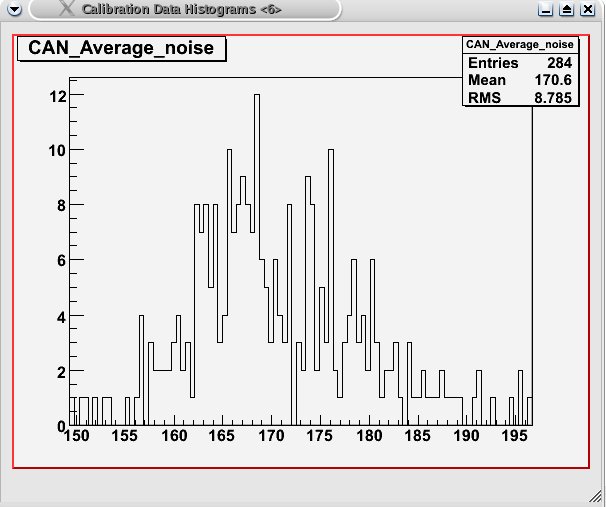}
\end{minipage}\hfill
\caption{Left: result of the Pixel threshold scan. The double peak structure around the 4000 electrons threshold is due to the difference in temperatures between the calibration (during assembly on various surface sites) and the scan (with the final cooling during the commissioning). Right : result of the Pixel noise measurement. The noise is measured to $\sim$ 170 electrons, in good agreement with the expectation of $\sim$ 160 electrons.} \label{PixelFig}
\end{figure*}

A complete description of the {\bf Pixel}, of its electronics, and of the tuning of its timing parameters can be found in \cite{DDthesis}. The {\bf Pixel} detector commissioning has started on April 25$^{th}$, but was quickly interrupted by a cooling plant failure. This cooling plant uses 6 compressors. Three of them failed after few days of running, the fault being the magnetic coupling between the motor and the compressor shaft. In case of slippage during an extended period of time, eddy currents considerably heat the metallic parts, and damage the neighboring plastic parts. This problem is now fixed, but all the compressors have to be dismounted and cleaned, new filters have to be introduced in the cooling lines, and a chemical analysis of the polluted C$_3$F$_8$ coolant has to be performed. This introduces an unexpected delay in the {\bf Pixel} commissioning, as well as in the {\bf SCT} commissioning, as it is depending on the same evaporative cooling system \cite{SiliconCooling}.\\
Nevertheless, a fraction of the {\bf Pixel} modules have been tested in a 6 days run. 11/1744 of the modules are not operable at the present time, mostly due to failing High Voltage connections. 87/88 of the cooling loops have been operated : 3 are leaky, and 1 is probably not recoverable. The three problematic cooling loops would probably not be operated on this first year of data taking, until the exact implication of a C$_3$F$_8$ pollution in the {\bf ID} environment is understood. 87\% of the modules have been tested for the cooling sign-off, which means powered, and tested for communication and configuration. Threshold scans have been performed on 60\% of them. The tuning of the timing parameter in the Back of Crate (see \cite{DDthesis}) has been done only for some of the failing modules, and the commissioning will restart as soon as the cooling unit is totally commissionned and the beam pipe bake-out done (the Beryllium beam pipe has to be bake-out at 200$^o$C to activate the ion pump, to reach the requested vacuum during the {\bf LHC} operations). \\
The threshold scans performed show that the thresholds are set around ~4000 electrons. Variations {\it wrt} this value are expected, as the calibration was made during the module assembly on the surface at a different temperature than the one during the commissioning. The noise is measured at 170 electrons, as expected (see fig. \ref{PixelFig}). This number has to be compared to the ~$\sim$ 20000 electrons corresponding to the most probable energy loss for a minimum ionizing particle in a 250 $\mu m$ thick silicon sensor. {\bf Pixel} detector is basically a noise free detector. \\
The {\bf Pixel} has also implemented a {\bf ROD} simulator, in order to join the data acquisition chain in the various ATLAS tests, even if the detector can not be powered due to the cooling plant failure. More on the {\bf Pixel} detector commissioning can be found in \cite{JFArguin}.

%\begin{tabular}{ccc}
%\includegraphics[width=50mm]{threshold_pixel.jpg} & & \includegraphics[width=50mm]{noise_pixel.jpg} \\
%\end{tabular}

\section{CONCLUSIONS}
The ATLAS {\bf ID} commissioning is late, because of the cooling plant failure in the case of the {\bf Pixel} and {\bf SCT}, and because of the late {\bf ROD} delivery in the case of the {\bf TRT}. Nevertheless, the ATLAS {\bf ID} is ready on time for the first {\bf LHC} beams. All the {\bf TRT RODs} have been delivered at CERN and are presently under installation and tests. The cooling plant has been repaired and slightly modified (there is now a slippage detector) as well as the cooling system, where more filters have been introduced to avoid that any pollution of the coolant goes inside the detectors. \\
The {\bf TRT} is routinely running since months without any major problems. The plans until the first beam is to operate higher and higher fraction of the detector, following the {\bf RODs} insertion. Also, few more monitoring tools are expected, mainly to monitor the back-end electronics. \\
The {\bf SCT} and the {\bf Pixel} have to fix some opto-link failures. These failures are located at the back of crate level in the back-end electronics, so these problems can be fixed even in the absence of the cooling, and the replacement of the failing components is on-going. \\
The {\bf SCT} first participation to a cosmic test has been a success, and there are no major problems. It is waiting for the cooling to continue the detector commissioning, one of the goal being a high acquisition rate (100 kHz) test. \\
The {\bf Pixel} do not show any major problems, but most of the commissioning work is still to be done. The cooling is back and running. During the time of the conference, the bake-out of the beam pipe has been sucessfully performed. {\bf Pixel} commissioning has then to be finished before the first {\bf LHC} beams \cite{LE}.

\end{document}